# Does Environmental Economics lead to patentable research?

[1]Xiaojun Hu, [2,3]Ronald Rousseau, [4]Sandra Rousseau

[1]Medical Information Center and Department of Neurology of Affiliated Hospital 2, Zhejiang University School of Medicine, Hangzhou 310058 (China) ;

xjhu@zju.edu.cn

[2, 3] KU Leuven, MSI, Facultair Onderzoekscentrum ECOOM, Naamsestraat 61, 3000 Leuven, (Belgium);

ronald.rousseau@kuleuven.be

&

University of Antwerp, Faculty of Social Sciences, Middelheimlaan 1, 2020 Antwerpen, Belgium;

ronald.rousseau@uantwerpen.be

[4] KU Leuven, CEDON, B-1000 Brussel (Belgium);

sandra.rousseau@kuleuven.be

**Abstract.** In this feasibility study, the impact of academic research from social sciences and humanities on technological innovation is explored through a study of citations patterns of journal articles in patents. Specifically we focus on citations of journals from the field of environmental economics in patents included in an American patent database (USPTO). Three decades of patents have led to a small set of journal articles (85) that are being cited from the field of environmental economics. While this route of measuring how academic research is validated through its role in stimulating technological progress may be rather limited (based on this first exploration), it may still point to a valuable and interesting topic for further research.

## 1. Purpose of this investigation

Patents are generally related to fields such as material sciences, mechanics, computer technology, biotechnology, pharmacy and other 'hard' fields of science. This leads to the question: what is the contribution of social sciences and humanities to the intellectual property system as covered by patents. In this contribution we focus on a small subfield, namely environmental economics (Turner, Pearce &

Bateman, 1994) and in particular to the role played by journals as non-patent references. Our work can be seen as a partial validation study of academic research, in particular of environmental economics.

This work is in part inspired by a similar publication by Halevi & Moed (2012) which focused on the use of journals from the field of Library Science in patents. They used TotalPatent[TM] as their data source. Their most important conclusion was that it was Library Science research that informed and inspired the development of information retrieval solutions, sometimes years before the technology was available to translate these ideas into technical devices or computer algorithms. Of an initial list of 42 library journals, 8 were found to be cited in patents covered by TotalPatent™. In terms of journals the *Journal of the American Society for Information Science and Technology* was the highest cited with a total of 76 citations overall and 24 unique articles cited.

## 2. Introduction: what is Environmental Economics?

According to Kolstad (2000), environmental economics is defined as follows:

> Environmental economics is concerned with the impact of the economy on the environment, the significance of the environment on the economy, and the appropriate way of regulating economic activity so that balance is achieved among environmental, economic, and other social goals (With the economy being defined by producer behavior and consumers' desires).

The field's increasing importance for society and within science is testified by the fact that environmental economist William Nordhaus, received (half of) the Sveriges Riksbank Prize in Economic Sciences in Memory of Alfred Nobel (popularly known as the Nobel Prize for Economics) in 2018 "for integrating climate change into long-run macroeconomic analysis".

## 3. Methods: journals and used database

In this study we operationalize the field of environmental economics by a set of leading journals. As there is no Subject Category called Environmental Economics in the WoS/JCR, we take the journals used by Sandra Rousseau in (Rousseau, 2008). [Also: Rousseau et al., 2009] These are shown in Table 1. The first column gives the name of the journal and the date it was either established or changed name; if it changed name (this is necessary to know when searching for patents that cite this journal) the former name and the date the journal was established is shown in the second column. Finally, the third column gives the journal's WoS category in 2017. Ten of the eleven journals belong, among others, to *Economics*; eight belong to *Environmental Studies*. All journals are multidisciplinary in the sense that they belong to more than one WoS category.

Table 1. Journals used in this investigation, their former title (if applicable) and the WoS categories to which they belong.

| Journals (established/since) | Formerly (established) | WoS Categories (2017) |
|---|---|---|
| American Journal of Agricultural Economics (1968) | Journal of Farm Economics (1919) | Economics (SSCI) – Agricultural Economics and Policy (SCIE) |
| Australian Journal of Agricultural and Resource Economics (1997) | Australian Journal of Agricultural Economics (1957) | Economics (SSCI) – Agricultural Economics and Policy (SCIE) |
| Ecological Economics (1989) | | Ecology (SCIE) – Environmental Sciences (SCIE) – Economics (SSCI) – Environmental Studies (SSCI) |
| Energy Journal (1980) | | Energy & Fuels (SCIE) – Economics (SSCI) – Environmental Studies (SSCI) |
| Environment & Development Economics (1996) | | Economics (SSCI) – Environmental studies (SSCI) |
| Environmental & Resource Economics (1991) | | Economics (SSCI) – Environmental studies (SSCI) |
| Journal of Agricultural and Resource Economics (1992) | Western Journal of Agricultural Economics (1977) | Economics (SSCI) – Agricultural Economics & Policy (SCIE) |
| Journal of Environmental Economics and Management (1974) | | Business (SSCI) – Economics (SSCI) – Environmental Studies (SSCI) |
| Land Economics (1948) | The Journal of Land & Public Utility Economics (1925) | Economics (SSCI) – Environmental Studies (SSCI) |
| Natural Resources Journal (1961) | | Environmental Studies (SSCI) – Law (SSCI) |
| Resource and Energy Economics (1993) | Resources and Energy (1978) | Economics (SSCI) – Environmental Studies (SSCI) |

In this investigation we restrict patents to those included in the USPTO database, the database of the United States Patent and Trademark Office. Data were collected in December 2018.

## 4. Searching in the USPTO

We used Advanced Search in the USPTO PatFT subbase, which contains full text patent documents since 1976. Advanced search includes the search prefix OREF, which restricts queries to non-patent references (**O**ther **REF**erences). USPTO further allows phrase searching by using quotation marks. As it, moreover, uses a list of stop words (http://patft.uspto.gov/netahtml/PTO/help/stopword.htm) a search for e.g. the Australian Journal of Agricultural and Resource Economics is done using the query

OREF/"Australian Journal Agricultural Resource Economics".

where stop words were removed. Yet, in order to cover this journal's earlier title (see Table 1) we actually used the query:

OREF/"Australian Journal Agricultural"

Each query performed in this way led usually to many false positives. Hence we checked each result manually. We further tried the standard abbreviations for each journal mostly without any usable result. For instance, a search for OREF/"j environ econ manag" and OREF/"environ econ manag", both resulted in zero patents, while a search for JEEM (aiming for Journal of Environmental Economics and Management) resulted in one patent citing JEEM = Journal of Embryology and Experimental Morphology (when checking the cited article).

We further recall that, contrary to articles, patents often have the same title and largely the same content. They may differ only in the claim.

## 5. Results on journal level

Table 2 shows that this set of journals dealing with environmental economics received a total of 195 citations among which 85 different articles. Two journals did not contribute any patent: Natural Resources Journal and Environment & Development Economics. The first one is published by the University of New Mexico School of Law, while the second one is published by the Cambridge University Press. We have no idea why these journals were not cited. We can only point out that a journal published by a school of law may be less popular among inventors, while a journal aiming at research from both developing and developed countries may be less attractive among American inventors submitting to the USPTO and its examiners.

Table 2. Number of citations for each journal and number of different cited articles

| Journals | Number of citations | Different cited articles |
|---|---|---|
| American Journal of Agricultural Economics | 33 | 20 |
| Australian Journal of Agricultural and Resource Economics | 12 | 2 |
| Ecological Economics | 20 | 11 |
| Energy Journal | 39 | 19 |
| Environment & Development Economics | 0 | 0 |
| Environmental & Resource Economics | 30 | 5 |
| Journal of Agricultural and Resource Economics | 19 | 8 |
| Journal of Environmental Economics and Management | 12 | 8 |
| Land Economics | 27 | 12 |
| Natural Resources Journal | 0 | 0 |
| Resource and Energy Economics | 3 | 3 |
| TOTAL | 195 | 85 |

Even more than for journal article citations, non-patent citations often have re-citations, i.e. the same author citing the same article (White, 2000, 2001). An extreme case is the Australian Journal of Agricultural and Resource Economics which has only two cited articles: one which is cited nine times by an inventor and another which is cited three times by (another) inventor.

After removing the two journals that were never cited, we calculated Spearman rank correlations between age of the journal and the number of received citations and between age of the journal and different cited articles. These are equal to -0.05 and 0.25, respectively. This shows that the age of these journals does not explain received citations.

Next we calculated the average and median time (in years) between submission of the patent application and publication of the cited article. We also determined the average and median time (in months) between patent application and patent assignment. Data are shown in Table 3. Column two shows that for most journals the average time between publication of the cited article and its use in a patent application is between nine and ten years. Yet, the overall average is about 14 years. For most journals there is not much difference between average and median cited age (and remember that for the Australian journal, data refer to only two different articles). This is an indication that distributions are not highly skewed or have few outliers. The last column gives an indication about the time between submission of a patent application and finally receiving the patent (about four years).

Table 3. Average and median time between publications and their use in a patent application; average and median time between filing and being assigned the patent.

| Journals | Average (and median) age in years of cited articles | Average (and median) time in months before acceptance |
|---|---|---|
| American Journal of Agricultural Economics | 15.9 (14) | 49.3 (40) |
| Australian Journal of Agricultural and Resource Economics | 28.7 (36) | 52.1 (47) |
| Ecological Economics | 7.9 (8) | 48.4 (38.5) |
| Energy Journal | 9.8 (9) | 56.7 (55) |
| Environmental & Resource Economics | 9.4 (9.5) | 30.5 (31) |
| Journal of Agricultural and Resource Economics | 9.6 (9) | 35.3 (30) |
| Journal of Environmental Economics and Management | 9.7 (9) | 50.1 (40) |
| Land Economics | 28.2 (26) | 55.8 (40) |
| Resource and Energy Economics | 9.3 (7) | 46 (49) |
| ALL | 14.3 | 47.5 |

As to the time between filing the patent application and finally receiving it, the record among the patents included in our search goes to patent 8,458,082 which was assigned after 137 months (almost 12 years) to Interthinx, Inc. in California (inventors were: Halper, Wilson & Hourigan). The patent dealt with an automated loan risk assessment system. It did cite two environmental economics articles though.

On the other hand, Leland Leachman from Leachman Cattle of Colorado, needed only wait for four months before patent 8,725,557 was assigned to him. The explanation for this extremely short delay is that this patent is a continuation of an earlier patent application. We recall that a continuation application is filed when one has filed for a patent containing a set of claims and then one wants to add some more claims for the same invention at a later stage. This is possible, but the continuation application has to be filed before the earlier application is either abandoned or patented. In this case the original patent application and the continuation both cited (Williams et al., 2012).

## 6. Results on article level

*The oldest cited articles*

The oldest cited article dates from 1929. It was cited in 2009 and published in the American Journal of Agricultural Economics (Hardie & Strand, 1929). The second oldest one dates from 1936 and was published in the Journal of Land & Public Utility Economics (Burton, 1936). This article has been cited twice by the same group of inventors working for PriceLock Finance in Nebraska.

*Special cases: Cited articles that were published after the patent had been filed.*

We found two such cases: one was just a misprint where the publication age was filed as 2011 while it had to be 2001. The other one was more mysterious. The article was published in 2014, while the patent was filed in 2012. Given that citation was included by the applicant, is an inventor self-citation and that the inventors published an article with the same title in a conference proceedings this leads us to two possible explanations: or the inventors knew already the journal, exact volume (year) and page in which their proceedings article would be published, or the inventors were able to change the submitted file.

*Highest cited environmental economics articles in patents*

The notion of "highly-cited" environmental economics articles in patents turned out to be very relative, especially as the few cases of (relative) high numbers of citations in patents result from re-citations. Table 4 shows the most-cited articles (5 or more, with one exception because this article was cited by 3 different inventor groups). To put this in perspective we include the number of received citations in the WoS, compared to the number of citations received by the most-cited article in the same journal and same publication year (column 4) in December 2018 and the rank of this article among all publications of article or review type in the same journal and publication year (column 5), again in December 2018.

Table 4. Most-cited articles in patents

| Article | Received patent citations | Different investigator groups | Received citations – most cited article | Rank among publications of article type in same journal and for same publication year |
|---|---|---|---|---|
|  |  |  |  |  |

| | | | |
|---|---|---|---|
| Ahlheim, M. et al. (2002). Allowing for Household Preferences in Emission Trading - A Contribution to the Climate Policy Debate. Environmental and Resource Economics, 21(4), 317-34. | 26 | 1 | 11 - 443 | 41/57 |
| Wood, S. (1976). Combining Forecasts to Predict Property Values for Single-Family Residences. Land Economics, 52(2), 221-229. | 9 | 5 | 2 - 185 | 22/29 |
| Hardaker et al. (1973). Assesment of the output of a stochastic decision model. Australian Journal of Agricultural Economics, 17(3), 170-178. | 9 | 1 | 3 - 12 | 5/17 |
| Bailey, D.V. et al. (1995). , et al., Identifying Buyer Market Areas and the Impact of Buyer Concentration in Feeder Cattle Markets Using Mapping and Spatial Statistics. American Journal of Agricultural Economics, 77(2), 309-318. | 8 | 2 | 13 - 384 | 61/133 |
| Pritchett, J. G. et al. (1998). Optimal Choice of Generic Milk Advertising Expenditures by Media Outlet. Journal of Agricultural and Resource Economics, 23(1), 155-169. | 7 | 3 | 8 - 71 | 16/36 |
| Matsukawa, I. et al. (2000). Household Response to Incentive Payment for Load Shifting: A Japanese Time-of-Day Electricity Pricing Experiment. Energy Journal, 21(1), 73-86. | 6 | 5 | 10 - 137 | 15/23 |
| Solomon B.D. (1999). New directions in emissions the potential contribution of new institutional economics. Ecological Economics, 30(3), 371-387. | 6 | 1 | 24 – 879 | 60/106 |
| Smith, V.K. et al. (1998). Buying time: Real and hypothetical offers. Journal of Environmental Economics and Management, 36(3), 209-224. | 5 | 1 | 31 - 355 | 16/34 |
| Chavas et al. (1985). Modeling dynamic agricultural production response - The case of swine production. American Journal of Agricultural Economics, 67(3), 636-646. | 4 | 3 | 30 - 125 | 17/145 |

Only one article, namely that by Chavas et al. (1985), belongs to the first quartile among those articles (or reviews) in the same journal and the same publication year. Clearly, among this small set of articles, being cited in a patent bears no relation with the number of academic citations.

## 7. Discussion and conclusion

This short contribution shows that it is feasible to search for articles originating from a set of journals such as the environmental economics journals in this case. Yet, the USPTO is not an optimal choice as database. A broader database in which better queries are possible is needed. One point is that USPTO does not have a standardized form for journals, leading to many false positives. After we submitted this text for the ISSI 2019 conference Bryan et al. (2019) published a user's guide for in-text patent citations and a corresponding database. Yet, they did not cover any of the environmental economics journals, studied by us. For these reasons we intend to use The Lens (https://www.lens.org/ ) or Dimensions (https://www.dimensions.ai) in further investigations.

Our results suggest that in next step, a roadmap could be developed to trace the boundary function (between science and technology) of a journal or of a subfield. Although the numbers of articles involved could be rather low, this should not be a pretext to overlook their contribution (Hu & Rousseau, 2018; Tijssen, 2010; van Raan, 2017).

Another potential application is that, when performing more and broader investigations, a pattern of knowledge transition between science and technology, as shown by non-patent citations may become discernable. A dual investigation about citations of patents in academic journals may complete a map of knowledge transition between science and technology.

## 8. Further research questions

To finish this work in progress we mention some research questions that deserve further attention.

What is the relation (on journal level) between the journal's synchronous impact factor (different periods could be considered) and the (relative) number of patent citations. We expect though that a correlation would be low as the indicators measure something different.

Would making a distinction between citations by the applicant and by the examiner lead to different results?

Surely, for a complete investigation a better proxy for the field of environmental economics is necessary. If it would be possible to describe the field on an article basis, instead on a journal basis, that would certainly lead to a better result.

What are the major factors affecting the pattern of citation links between an article and a patent? Does geographical distance influence the citation behavior of inventors and examiners?

Are patents citing contributions in environmental economics related to the economy in general, or do they contribute to a "green economy"?

Can journal citations in patents help to measure the impact of research performed at academic institutions?

Clearly, for a thorough investigation a large database, not just the USPTO, is necessary.

Finally, whatever the topic of a patent investigation, it could be discussed in view of the struggle for scientific and technological leadership between the USA, Europe and China. Indeed, in 2017 China moved already into the second position (country level) as a source of international patent applications filed via WIPO (World Intellectual Property Organization) https://www3.wipo.int/ipstats.